\documentclass[twocolumn, prl, aps, superscriptaddress, longbibliography,showpacs,amsmath,amssymb,floatfix]{revtex4-1}
\usepackage{changes}
\usepackage{graphicx}% Include figure files
\usepackage{dcolumn}% Align table columns on decimal point
\usepackage{bm}% bold math
\usepackage{graphicx}                                   
\usepackage{amssymb}

\usepackage{amsmath}
\usepackage{epsfig}
\usepackage{xcolor}
\usepackage{tabu}
\usepackage{mathtools}
\usepackage[colorlinks,linkcolor=blue,anchorcolor=blue,citecolor=blue,urlcolor=blue]{hyperref}
\usepackage{physics}
\usepackage{float}
\usepackage{diagbox}
\usepackage{inputenc}

\begin{document}
    
\title{Plateau Gaps of Poisson Correctors Encode Metastable Reaction Rates}
% Validity and Breakdown of Additive Decomposition for Diffusion in a Random Potential
\author{Sang Yang}
\email{yangsang@mail.ustc.edu.cn}
%\affiliation{University of Science and Technology of China, Hefei, 230026, China}
\affiliation{Key Laboratory of Quantum Information, University of Science and Technology of China, Hefei 230026, China}
\affiliation{Hefei National Laboratory, University of Science and Technology of China, Hefei 230088, China}
\author{Zhixin Peng}
\email{yangsang@mail.ustc.edu.cn}
\affiliation{University of Science and Technology of China, Hefei, 230026, China}

%\author{Ming Gong}
%\email{gongm@ustc.edu.cn}
%\affiliation{Key Laboratory of Quantum Information, University of Science and Technology of China, Hefei 230026, China}
%\affiliation{Hefei National Laboratory, University of Science and Technology of China, Hefei 230088, China}
%\affiliation{Synergetic Innovation Center of Quantum Information and Quantum Physics, University of Science and Technology of China, Hefei 230026, China}

\date{\today}
	
\begin{abstract}
Metastable reaction rates are commonly inferred from transition-state fluxes, mean first-passage times, or fitted kinetic models. We show that they are directly encoded in the plateau gap of an occupation-time Poisson corrector. For a centered basin-occupation observable, the Poisson corrector develops metastable plateaus in the reactant and product basins, and their separation determines the forward and backward transition rates. This construction requires only the generator, stationary measure, and metastable partition, and therefore does not rely on a predefined transition-state surface. In overdamped and underdamped double-well dynamics, the plateau-gap rate recovers the Kramers, Grote-Hynes, and Pollak-Grabert-Hänggi hierarchy. The same corrector-martingale decomposition yields a reactive-noise density, revealing where stochastic forcing contributes to transitions in configuration or phase space. Thus, reaction rates and their fluctuation sources emerge from a single corrector field.

\end{abstract}
\maketitle

{\it Introduction.} Reaction-rate theory has long revolved around a deceptively simple question: when does a crossing become a reaction? Transition-state theory gives the earliest answer. It estimates the rate from the equilibrium flux through a dividing surface and assumes that every crossing is reactive. This picture is elegant, but it is also fragile: a trajectory may cross the nominal transition state and return.

Kramers transformed this static picture into a stochastic dynamical one by embedding barrier crossing in Brownian motion\cite{mel1991kramers}. In the high-friction regime, escape is limited by spatial diffusion over the barrier; in the low-friction regime, it is limited by energy diffusion inside the well. The resulting turnover showed that the heat bath is not merely a passive thermostat, but an active dynamical agent controlling the reaction. Grote and Hynes then sharpened the meaning of a reactive crossing in underdamped dynamics: the relevant coordinate near the barrier is not the configuration coordinate alone, but an unstable phase-space mode. Pollak, Grabert, and Hänggi completed this hierarchy by combining inertial recrossing with energy-loss effects, showing that even after a barrier crossing, a trajectory may fail to depopulate the reactant region.

This development progressively refined the meaning of a reactive event. Yet the final output is still usually a scalar rate constant. The global slow mode whose relaxation defines the rate is not directly displayed, and the region where stochastic forcing generates reactive fluctuations remains hidden. This limitation becomes more severe in underdamped or nonequilibrium metastable systems, where a useful dividing surface, scalar potential, or conventional transition-state surface may not be available.

Here we take a different route. Rather than starting from a dividing surface and correcting for recrossing, we start from the occupation time of a metastable basin. For a Markov process with generator $\mathcal{L}$, stationary measure $\mu$, and metastable sets $A$ and $B$, we define the centered basin observable
\begin{equation}
h=\mathbf{1}_B-\mu_B,
\end{equation}
and solve the Poisson equation
\begin{equation}
\mathcal{L} \chi=-h .
\end{equation}
In a metastable regime, the corrector $\chi$ develops two plateaus in the reactant and product basins. We show that their separation directly gives the forward and backward transition rates. Thus the rate is not introduced through a guessed transition state; it is read from the plateau structure of the global occupation-time response.

The same construction also provides information beyond scalar rates. The additive-functional decomposition
\begin{equation}
\int_0^t h\left(Z_s\right) d s=\chi\left(Z_0\right)-\chi\left(Z_t\right)+M_t
\end{equation}
defines a martingale whose quadratic variation gives a reactive-noise density. This density identifies where stochastic forcing contributes to transitions in configuration or phase space. Applied to double-well dynamics, the plateau-gap rate recovers the Kramers, Grote-Hynes, and Pollak-Grabert-Hänggi hierarchy, while the martingale density exposes the fluctuation structure hidden behind these classical rate formulas.

{\it Corrector-platform rate formula.} We first formulate the rate extraction principle in a general Markov setting. Let $Z_t$ be a stationary Markov process with generator $\mathcal{L}$ and invariant measure $\mu$. Let $A$ and $B$ be two metastable sets, interpreted as reactant and product basins. We define the centered occupation observable of the product basin,
\begin{equation}
h(z)=\mathbf{1}_B(z)-\mu_B, \quad \mu_B=\int \mathbf{1}_B(z) \mu(d z)
\end{equation}
and introduce the Poisson corrector $\chi$ by
\begin{equation}
\mathcal{L} \chi(z)=-h(z), \quad \int \chi(z) \mu(d z)=0 .
\end{equation}
For the additive occupation-time fluctuation,
\begin{equation}
A_t=\int_0^t h\left(Z_s\right) d s
\end{equation}
Dynkin's formula gives the exact decomposition
\begin{equation}
A_t=\chi\left(Z_0\right)-\chi\left(Z_t\right)+M_t,
\end{equation}
where $M_t$ is a martingale. This decomposition separates the occupation-time fluctuation into a corrector term, which depends only on the endpoints, and a martingale term, which accumulates stochastic forcing along the trajectory.

\begin{figure}
    \centering
    \includegraphics[width=1\linewidth]{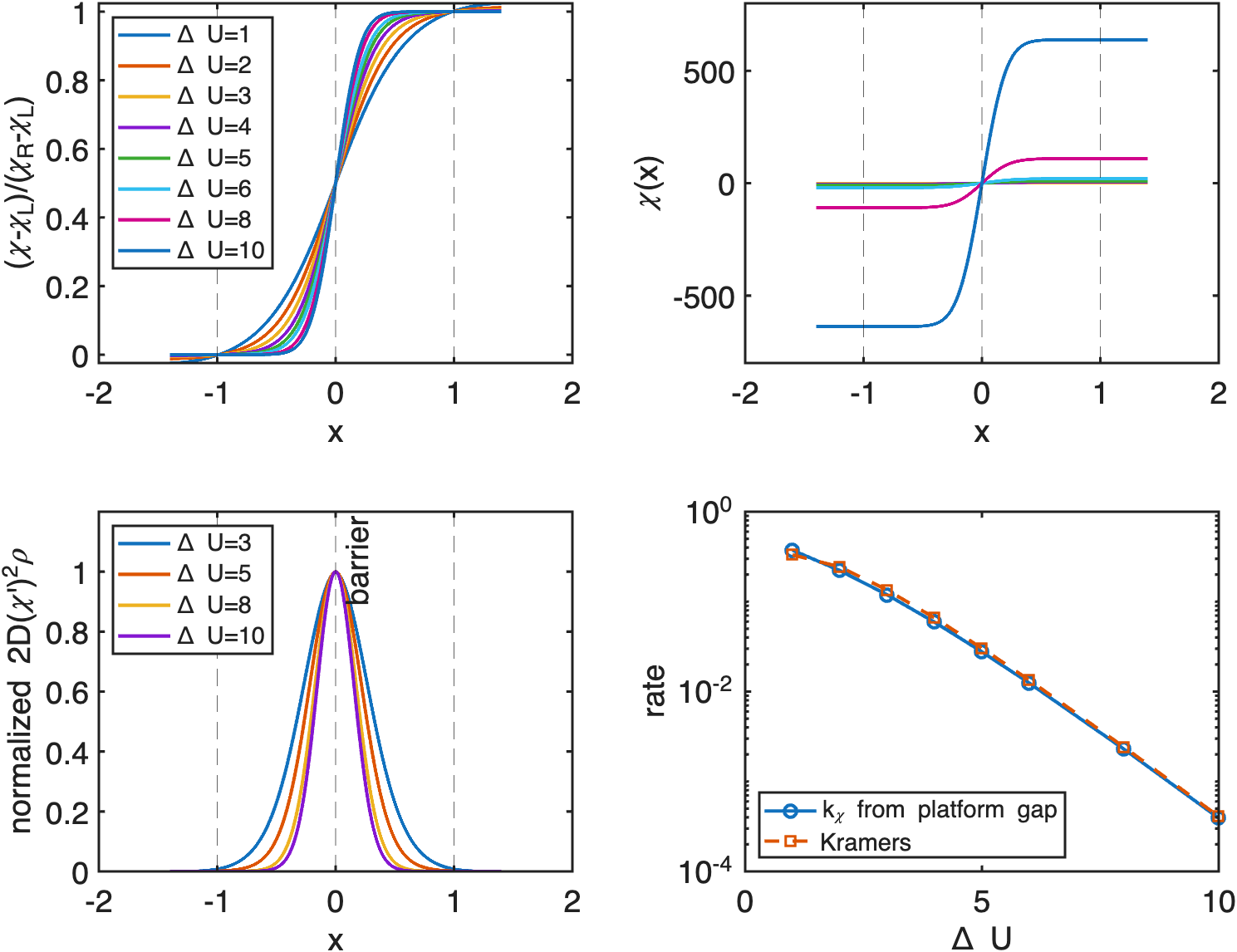}
    \caption{Overdamped corrector plateaus and recovery of the Kramers rate.(a) Normalized Poisson corrector $\left(\chi-\chi_L\right) /\left(\chi_R-\chi_L\right)$ for the overdamped quartic double well $U(x)=\Delta U\left(x^2-1\right)^2$ at different barrier heights. As $\Delta U$ increases, the corrector develops two welldefined plateaus in the left and right wells, with a sharp transition layer near the barrier. (b) Raw corrector $\chi(x)$, showing that the plateau gap $\chi_R-\chi_L$ grows rapidly with barrier height. (c) Normalized martingale density $2 D\left[\chi^{\prime}(x)\right]^2 \rho_{\text {eq }}(x)$, which localizes near the barrier, indicating that long-time occupation fluctuations are generated by rare interwell transitions rather than intrawell thermal fluctuations. (d) Escape rate extracted from the plateau gap, $k_\chi=1 /\left[2\left(\chi_R-\chi_L\right)\right]$, compared with the overdamped Kramers asymptote. The agreement confirms that the Kramers time is encoded in the corrector plateau gap.}
    \label{fig-fig2}
\end{figure}

The crucial observation is that, in a two-state metastable regime, the solution $\chi$ becomes nearly constant inside each metastable basin. We write
\begin{equation}
\begin{array}{ll}
\chi(z) \simeq \chi_A, & z \in A, \\
\chi(z) \simeq \chi_B, & z \in B .
\end{array}
\end{equation}
The difference
\begin{equation}
\Delta \chi=\chi_B-\chi_A
\end{equation}
is the corrector plateau gap. We now show that this gap directly gives the transition rates. In the metastable limit, the slow dynamics is well approximated by a two-state Markov process,
\begin{equation}
A \underset{k_{B \rightarrow A}}{\stackrel{k_{A \rightarrow B}}{\rightleftharpoons}} B .
\end{equation}
For any function $g$ on the two states, the coarse-grained generator acts as
\begin{equation}
\begin{aligned}
\mathcal{L} g(A) & =k_{A \rightarrow B}[g(B)-g(A)], \\
\mathcal{L} g(B) & =k_{B \rightarrow A}[g(A)-g(B)] .
\end{aligned}
\end{equation}
Since the centered occupation observable satisfies
\begin{equation}
h(A)=-\mu_B, \quad h(B)=\mu_A, \quad \mu_A=1-\mu_B,
\end{equation}
the Poisson equation $\mathcal{L} \chi=-h$ gives, in state $A$,
\begin{equation}
k_{A \rightarrow B}\left(\chi_B-\chi_A\right)=\mu_B,
\end{equation}
and in state $B$,
\begin{equation}
k_{B \rightarrow A}\left(\chi_A-\chi_B\right)=-\mu_A .
\end{equation}
\begin{equation}
k_{A \rightarrow B}=\frac{\mu_B}{\Delta \chi}, \quad k_{B \rightarrow A}=\frac{\mu_A}{\Delta \chi} .
\end{equation}
For a symmetric two-state system, $\mu_A=\mu_B=1 / 2$, so the two rates are equal and
\begin{equation}
k=\frac{1}{2 \Delta \chi}
\end{equation}
This formula is the central result of the work: a metastable reaction rate can be extracted from the plateau gap of the occupation-time Poisson corrector. The same construction also identifies the fluctuation source. For a diffusion process
\begin{equation}
d Z_t=b\left(Z_t\right) d t+\sigma\left(Z_t\right) d W_t
\end{equation}
with diffusion matrix
\begin{equation}
a(z)=\sigma(z) \sigma(z)^T
\end{equation}
the martingale in the decomposition is
\begin{equation}
M_t=\int_0^t \nabla \chi\left(Z_s\right)^T \sigma\left(Z_s\right) d W_s
\end{equation}
Its quadratic variation is
\begin{equation}
\langle M\rangle_t=\int_0^t \nabla \chi\left(Z_s\right)^T a\left(Z_s\right) \nabla \chi\left(Z_s\right) d s
\end{equation}
Thus the local martingale density is
\begin{equation}
r(z)=\nabla \chi(z)^T a(z) \nabla \chi(z)
\end{equation}
The stationary weighted density
\begin{equation}
R(z)=r(z) \mu(z)
\end{equation}
measures which regions of state space contribute most strongly to long-time occupation fluctuations. Therefore, the corrector plateau gap gives the rate, while the martingale density gives the spatial or phase-space source of reactive noise.

\begin{figure}
    \centering
    \includegraphics[width=1\linewidth]{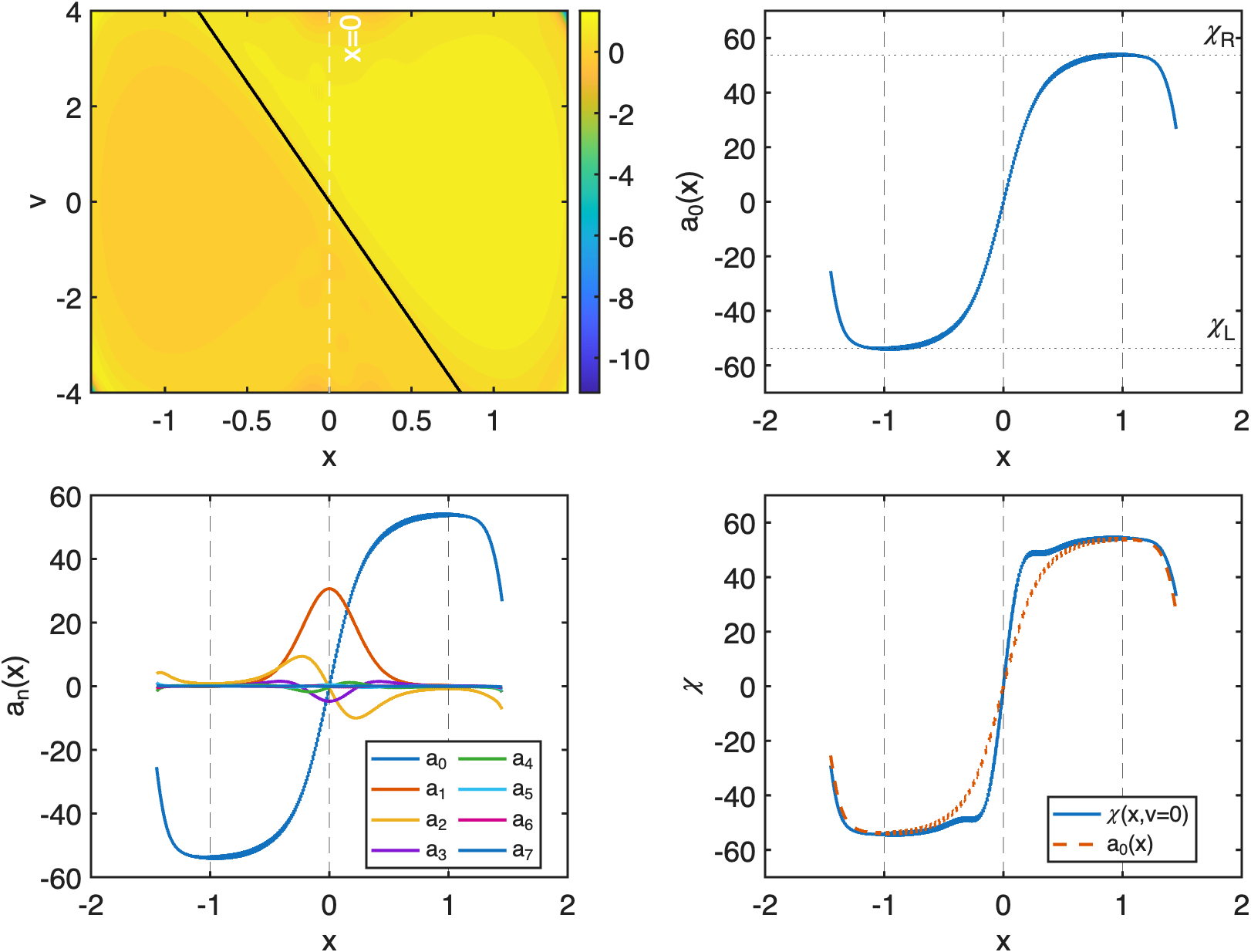}
    \caption{Underdamped corrector as a phase-space plateau field.(a) Hermite-Galerkin solution of the underdamped Poisson corrector $\chi(x, v)$ for the quartic double well. The white dashed line denotes the naive configurational dividing surface $x=0$, while the black line denotes the local unstable phase-space separatrix $v-\lambda_{-} x=0$. The corrector transition layer is tilted in phase space, showing that the reactive coordinate is not purely configurational. (b) Velocity-averaged corrector $a_0(x)$, which forms left- and right-basin plateaus $\chi_L$ and $\chi_R$. The plateau gap determines the rate $k_\chi=1 /\left[2\left(\chi_R-\chi_L\right)\right]$. (c) First Hermite coefficients $a_n(x)$ in the expansion $\chi(x, v)= \sum_n a_n(x) \mathrm{He}_n(\sqrt{\beta} v)$. The zeroth mode carries the dominant slow-platform structure, while higher modes encode velocity-space nonequilibrium and inertial corrections. (d) Comparison between the $v=0$ cut of the full corrector and the velocity-averaged mode $a_0(x)$, showing that $a_0$ captures the principal plateau gap while nonzero velocity modes reshape the transition layer.}
    \label{fig-fig3}
\end{figure}

This general result is independent of whether the dynamics is overdamped or underdamped, reversible or irreversible. It only assumes the existence of a stationary measure and a metastable partition. We now show how this abstract plateau-gap formula reduces to the classical Kramers result in a solvable overdamped double well, and then how it extends to underdamped phase-space dynamics.

{\it Overdamped Kramers limit.} Having established the plateau-gap formula in a general metastable setting, we first examine an overdamped double well, where the corrector can be written explicitly. Consider
\begin{equation}
d X_t=-D \beta U^{\prime}\left(X_t\right) d t+\sqrt{2 D} d W_t,
\end{equation}
with stationary density
\begin{equation}
\rho_{\mathrm{eq}}(x)=Z^{-1} e^{-\beta U(x)} .
\end{equation}
The generator is
\begin{equation}
\mathcal{L}=D\left(\partial_x^2-\beta U^{\prime}(x) \partial_x\right) .
\end{equation}
For the right-basin occupation observable,
\begin{equation}
h(x)=\mathbf{1}_R(x)-\mu_R,
\end{equation}
the Poisson equation
\begin{equation}
\mathcal{L} \chi=-h
\end{equation}
can be written in self-adjoint form as
\begin{equation}
D e^{\beta U(x)} \partial_x\left[e^{-\beta U(x)} \chi^{\prime}(x)\right]=-h(x) .
\end{equation}
Thus,
\begin{equation}
\chi^{\prime}(x)=-\frac{e^{\beta U(x)}}{D} \int_{-\infty}^x h(y) e^{-\beta U(y)} d y .
\end{equation}
This formula already reveals the plateau mechanism. In a high-barrier double well, the integral is dominated by the local equilibrium weight inside the wells, while the factor $e^{\beta U(x)}$ amplifies the barrier region. Consequently, $\chi^{\prime}(x)$ is strongly localized near the barrier, and $\chi(x)$ becomes nearly constant inside each basin. The corrector therefore develops two plateaus,
\begin{equation}
\begin{array}{ll}
\chi(x) \simeq \chi_L, & x \in L, \\
\chi(x) \simeq \chi_R, & x \in R .
\end{array}
\end{equation}
For a symmetric double well,
\begin{equation}
h(x)= \begin{cases}-1 / 2, & x \in L, \\ +1 / 2, & x \in R .\end{cases}
\end{equation}
Near the barrier $x_b$, the integral in $\chi^{\prime}(x)$ is dominated by the left well:
\begin{equation}
\int_{-\infty}^{x_b} h(y) e^{-\beta U(y)} d y \simeq-\frac{1}{2} Z_L,
\end{equation}
where
\begin{equation}
Z_L=\int_L e^{-\beta U(y)} d y
\end{equation}
Therefore, in the barrier layer,
\begin{equation}
\chi^{\prime}(x) \simeq \frac{Z_L}{2 D} e^{\beta U(x)} .
\end{equation}
The platform gap is then
\begin{equation}
\Delta \chi=\chi_R-\chi_L \simeq \frac{Z_L}{2 D} \int_{\text {barrier }} e^{\beta U(x)} d x .
\end{equation}
Using the plateau-gap formula,
\begin{equation}
k_\chi=\frac{1}{2 \Delta \chi}
\end{equation}
we obtain
\begin{equation}
k_\chi \simeq \frac{D}{Z_L \int_{\text {barrier }} e^{\beta U(x)} d x}
\end{equation}
Applying Laplace's method near the left minimum $x_a$ and the barrier top $x_b$,
\begin{equation}
\begin{aligned}
& U(x) \simeq U\left(x_a\right)+\frac{1}{2} U^{\prime \prime}\left(x_a\right)\left(x-x_a\right)^2, \\
& U(x) \simeq U\left(x_b\right)-\frac{1}{2}\left|U^{\prime \prime}\left(x_b\right)\right|\left(x-x_b\right)^2,
\end{aligned}
\end{equation}
gives
\begin{equation}
Z_L \simeq e^{-\beta U\left(x_a\right)} \sqrt{\frac{2 \pi}{\beta U^{\prime \prime}\left(x_a\right)}},
\end{equation}
and
\begin{equation}
\int_{\text {barrier }} e^{\beta U(x)} d x \simeq e^{\beta U\left(x_b\right)} \sqrt{\frac{2 \pi}{\beta\left|U^{\prime \prime}\left(x_b\right)\right|}}
\end{equation}
Thus,
\begin{equation}
k_\chi \simeq \frac{D \beta}{2 \pi} \sqrt{U^{\prime \prime}\left(x_a\right)\left|U^{\prime \prime}\left(x_b\right)\right|} e^{-\beta\left[U\left(x_b\right)-U\left(x_a\right)\right]} .
\end{equation}
Using
\begin{equation}
D=\frac{1}{\beta \gamma},
\end{equation}
we recover the overdamped Kramers rate,
\begin{equation}
k_{\mathrm{OD}}=\frac{\sqrt{U^{\prime \prime}\left(x_a\right)\left|U^{\prime \prime}\left(x_b\right)\right|}}{2 \pi \gamma} e^{-\beta \Delta U} .
\end{equation}
Therefore, in the overdamped high-barrier limit, the Kramers escape time is precisely encoded in the corrector plateau gap.

\begin{figure*}
    \centering
    \includegraphics[width=1\linewidth]{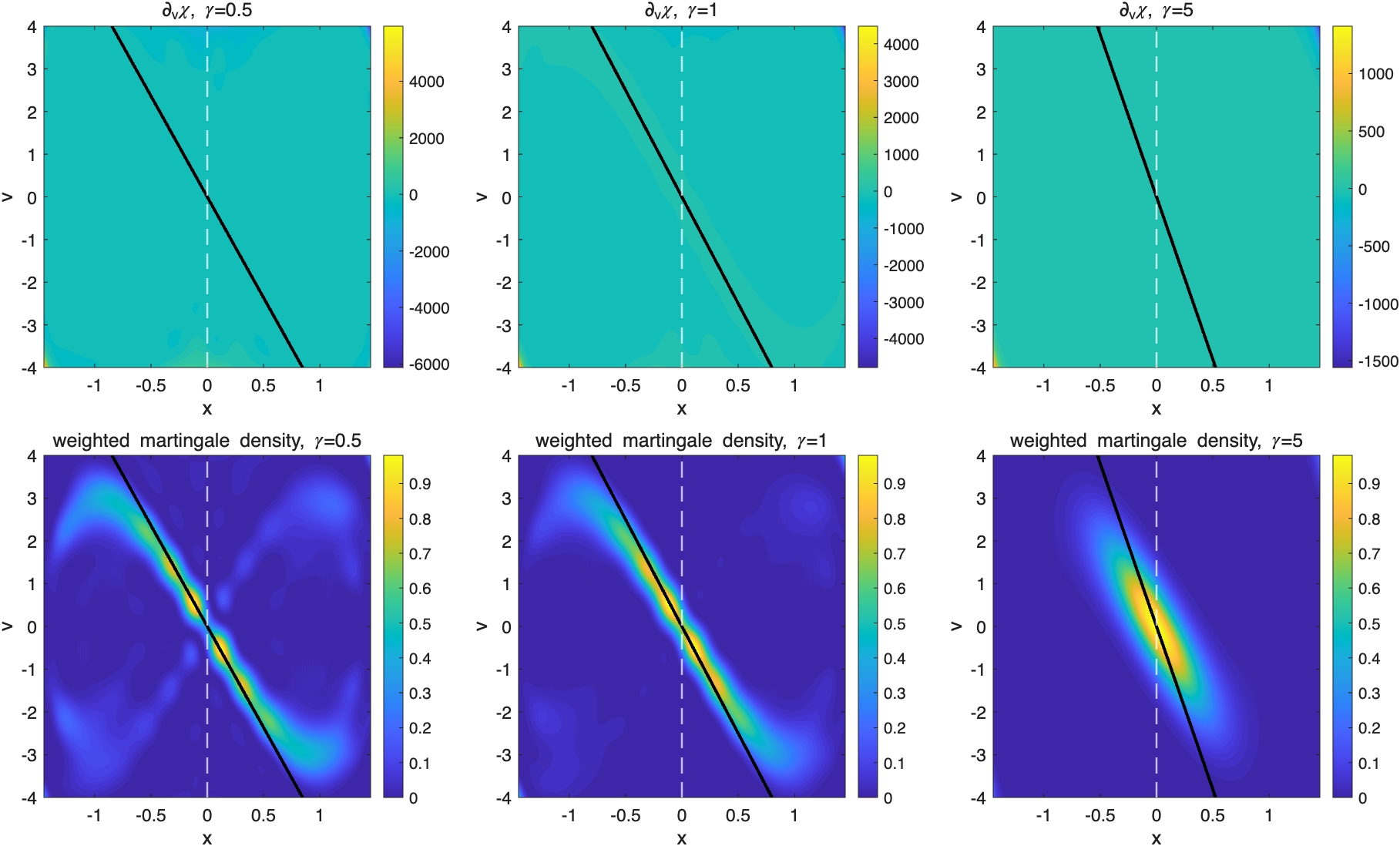}
    \caption{Martingale density reveals the phase-space source of reactive fluctuations.Top row: velocity derivative $\partial_v \chi(x, v)$ of the underdamped corrector for different friction coefficients $\gamma$. Since the thermal noise enters only through the velocity equation, $\partial_v \chi$ directly determines the martingale channel. Bottom row: normalized equilibrium-weighted martingale density $(2 \gamma / \beta)\left[\partial_v \chi(x, v)\right]^2 \rho_{\text {eq }}(x, v)$. The white dashed line marks the naive surface $x=0$, and the black line marks the local unstable separatrix $v-\lambda_{-} x=0$. The reactive-noise density is concentrated near the phase-space separatrix rather than the configurational barrier alone. As friction increases, the reactive-noise hotspot becomes more localized near the barrier region, consistent with the crossover toward overdamped Kramers behavior.}
    \label{fig-fig4}
\end{figure*}

For the quartic double well used in our numerical tests,
\begin{equation}
U(x)=\Delta U\left(x^2-1\right)^2,
\end{equation}
the minima are at
\begin{equation}
x_a= \pm 1,
\end{equation}
and the barrier is at
\begin{equation}
x_b=0 .
\end{equation}
The curvatures are
\begin{equation}
\begin{aligned}
U^{\prime \prime}( \pm 1) & =8 \Delta U \\
\left|U^{\prime \prime}(0)\right| & =4 \Delta U
\end{aligned}
\end{equation}
The Kramers rate becomes
\begin{equation}
k_{\mathrm{OD}}=\frac{2 \sqrt{2} \Delta U}{\pi \gamma} e^{-\beta \Delta U} .
\end{equation}
Numerically, the corrector computed from the integral formula forms increasingly sharp plateaus as $\Delta U$ increases. The rate extracted from
\begin{equation}
k_\chi=\frac{1}{2\left(\chi_R-\chi_L\right)}
\end{equation}
converges to $k_{\mathrm{OD}}$ in the high-barrier regime.
The same example also illustrates the martingale-density interpretation. For overdamped dynamics, the martingale part is
\begin{equation}
M_t=\sqrt{2 D} \int_0^t \chi^{\prime}\left(X_s\right) d W_s
\end{equation}
Its local quadratic-variation density is
\begin{equation}
r_{\mathrm{OD}}(x)=2 D\left[\chi^{\prime}(x)\right]^2 .
\end{equation}
The equilibrium-weighted contribution is
\begin{equation}
R_{\mathrm{OD}}(x)=2 D\left[\chi^{\prime}(x)\right]^2 \rho_{\mathrm{eq}}(x) .
\end{equation}
Although the stationary probability density is largest in the wells, $R_{\mathrm{OD}}(x)$ is localized near the barrier in the metastable regime. This shows that the long-time occupation fluctuation is generated mainly by rare reactive passages, not by intrawell thermal motion. Thus, the overdamped limit provides two pieces of intuition: the rate is the inverse corrector plateau gap, and the martingale density identifies the barrier region as the source of reactive noise.

{\it Underdamped phase-space corrector.} The overdamped case shows most transparently how the Kramers time appears as a corrector plateau gap. The more stringent test is underdamped dynamics, where the state is not a position $x$, but a phasespace point $(x, v)$. We consider
\begin{equation}
\begin{gathered}
d X_t=V_t d t \\
d V_t=\left[-\gamma V_t-U^{\prime}\left(X_t\right)\right] d t+\sqrt{\frac{2 \gamma}{\beta}} d W_t
\end{gathered}
\end{equation}
The generator is
\begin{equation}
\mathcal{L}=v \partial_x+\left[-\gamma v-U^{\prime}(x)\right] \partial_v+\frac{\gamma}{\beta} \partial_v^2
\end{equation}
For the same right-basin occupation observable,
\begin{equation}
h(x, v)=h(x)=\mathbf{1}_R(x)-\mu_R
\end{equation}
the corrector satisfies the hypoelliptic Poisson equation
\begin{equation}
v \partial_x \chi+\left[-\gamma v-U^{\prime}(x)\right] \partial_v \chi+\frac{\gamma}{\beta} \partial_v^2 \chi=-h(x)
\end{equation}
The central difference from the overdamped problem is that the corrector is now a phase-space field,
\begin{equation}
\chi=\chi(x, v)
\end{equation}
It cannot be reduced to a function of $x$ alone. Indeed, if one tries $\chi(x, v)=\chi_0(x)$, then
\begin{equation}
\mathcal{L} \chi_0=v \chi_0^{\prime}(x)
\end{equation}
which is odd in velocity, whereas the source term $-h(x)$ is even in velocity. Thus a purely configurational corrector cannot solve the underdamped Poisson equation. Velocity-dependent components are required.

To compute the global corrector, we expand it in Hermite polynomials of the thermal velocity variable,
To compute the global corrector, we expand it in Hermite polynomials of the thermal velocity variable,
\begin{equation}
\chi(x, v)=\sum_{n=0}^M a_n(x) \mathrm{He}_n(\sqrt{\beta} v) .
\end{equation}
This basis is natural because the equilibrium velocity distribution is Gaussian. Projecting the Poisson equation onto Hermite modes gives the coupled one-dimensional system
\begin{equation}
    \begin{aligned}
        &\frac{1}{\sqrt{\beta}} a_{m-1}^{\prime}(x)+(m+1)\left[\frac{1}{\sqrt{\beta}} a_{m+1}^{\prime}(x)\right.\\
        &\left.-\sqrt{\beta} U^{\prime}(x) a_{m+1}(x)\right]-\gamma m a_m(x)=-h_m(x),
    \end{aligned}
\end{equation}
with $a_{-1}=0$. For a position-only observable,
\begin{equation}
h_0(x)=h(x), \quad h_m(x)=0, \quad m \geq 1 .
\end{equation}
The numerical solution shows a clear separation of roles. The zeroth coefficient $a_0(x)$, which is the velocity-averaged corrector, carries the dominant two-plateau structure. It is this slow component whose platform gap determines the transition rate. The higher Hermite coefficients are smaller in amplitude, but they are not negligible errors. They encode the velocity-space non-equilibrium required to make the underdamped dynamics compatible with the even source $h(x)$. In physical terms, $a_0$ carries the occupation slow mode, while $a_{n \geq 1}$ encode inertial correction, recrossing, and the martingale channel.

\begin{figure}
    \centering
    \includegraphics[width=1\linewidth]{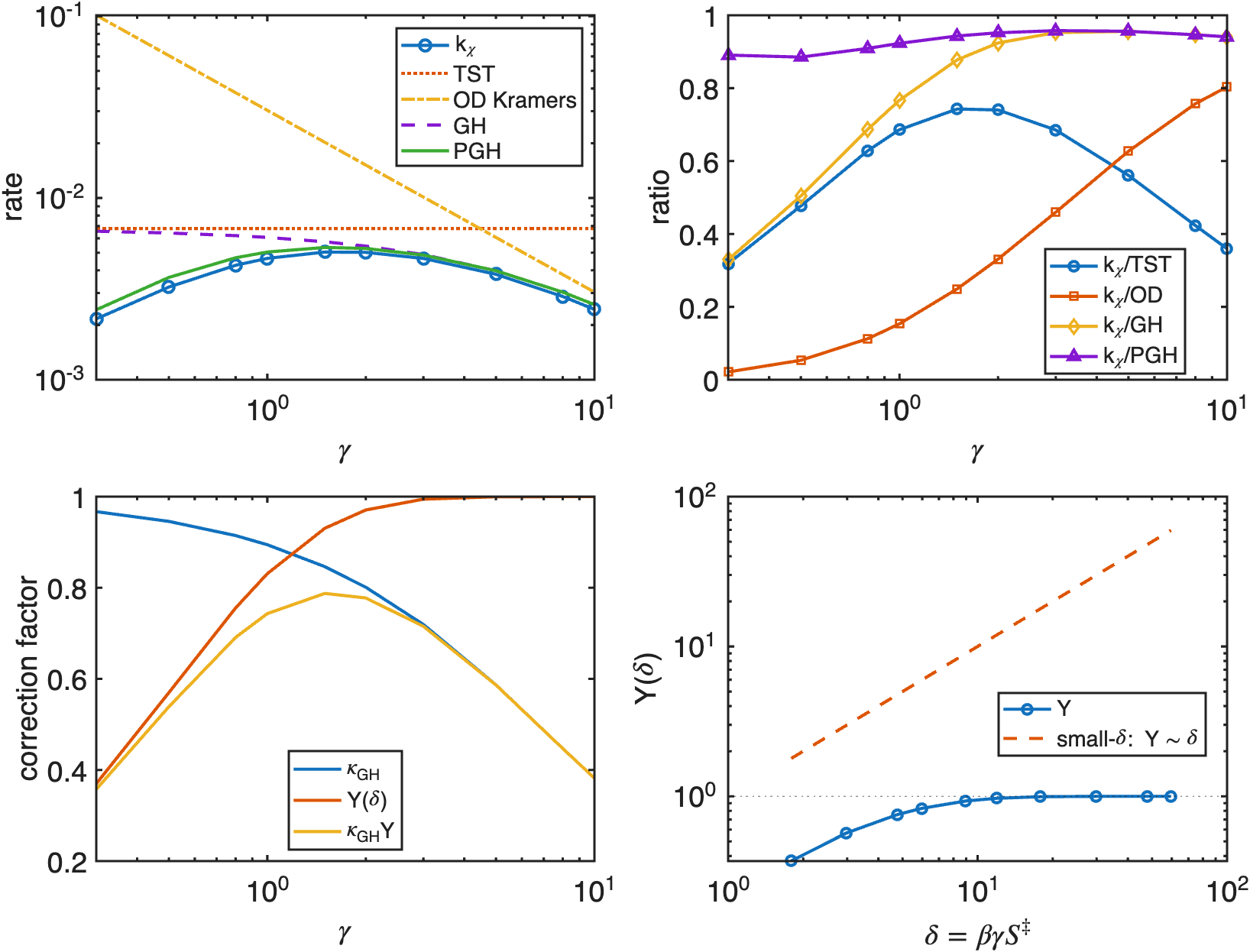}
    \caption{Plateau-gap rate across the Kramers-Grote-Hynes-PGH hierarchy.(a) Friction dependence of the rate extracted from the corrector plateau gap, $k_\chi$, compared with transition-state theory (TST), the overdamped Kramers asymptote, the Grote-Hynes (GH) rate, and the Pollak-Grabert-Hänggi (PGH) turnover expression. The plateau-gap rate follows the turnover trend and connects the moderate-friction and high-friction regimes. (b) Ratios of $k_\chi$ to the reference rates. The comparison shows that $k_\chi$ approaches the PGH/GH behavior over the underdamped regime and approaches the overdamped Kramers rate at large friction. (c) Correction factors $\kappa_{\mathrm{GH}}=\lambda_{+} / \omega_b, \Upsilon(\delta)$, and their product, illustrating the separate roles of phase-space recrossing and energy depopulation. (d) Depopulation factor $\Upsilon(\delta)$ as a function of the dimensionless energy-loss parameter $\delta=\beta \gamma S^{\ddagger}$, together with the small- $\delta$ asymptote. These results show that the corrector plateau gap provides a unified rate observable across the classical rate-theory hierarchy.}
    \label{fig-fig5}
\end{figure}

This structure is especially transparent in the high-friction limit. When $\gamma$ is large, the velocity relaxes rapidly and the underdamped corrector admits the expansion
\begin{equation}
\chi_{\mathrm{UD}}(x, v)=\chi_{\mathrm{OD}}(x)+\frac{v}{\gamma} \chi_{\mathrm{OD}}^{\prime}(x)+O\left(\gamma^{-2}\right),
\end{equation}
where $\chi_{\mathrm{OD}}(x)$ is the overdamped corrector. The first term gives the dominant configurational plateau, while the second term is the leading velocity correction. This expansion also confirms the consistency of the martingale densities. Since the underdamped noise acts only on the velocity coordinate,
\begin{equation}
M_t=\sqrt{\frac{2 \gamma}{\beta}} \int_0^t \partial_v \chi\left(X_s, V_s\right) d W_s
\end{equation}
we have
\begin{equation}
\partial_v \chi_{\mathrm{UD}} \simeq \frac{1}{\gamma} \chi_{\mathrm{OD}}^{\prime}(x) .
\end{equation}
Thus,
\begin{equation}
\frac{2 \gamma}{\beta}\left[\partial_v \chi_{\mathrm{UD}}\right]^2 \simeq \frac{2}{\beta \gamma}\left[\chi_{\mathrm{OD}}^{\prime}(x)\right]^2 .
\end{equation}
Using
\begin{equation}
D=\frac{1}{\beta \gamma},
\end{equation}
this becomes
\begin{equation}
\frac{2 \gamma}{\beta}\left[\partial_v \chi_{\mathrm{UD}}\right]^2 \simeq 2 D\left[\chi_{\mathrm{OD}}^{\prime}(x)\right]^2 .
\end{equation}
Therefore, in the high-friction limit, the underdamped martingale density reduces to the overdamped martingale density.

However, away from the overdamped limit, the corrector is genuinely phase-space dependent. The transition layer of $\chi(x, v)$ is tilted in phase space, reflecting the fact that crossing the barrier coordinate is not sufficient for reaction. A trajectory near the barrier may still return to the reactant basin depending on its velocity. The corrector plateau therefore encodes not only which side of the barrier the particle occupies, but also whether its phase-space direction belongs to a reactive or nonreactive sector.

This phase-space structure explains why the plateau-gap method naturally captures inertial recrossing. Near the barrier top, set $x_b=0$ and approximate
\begin{equation}
U(x) \simeq U_b-\frac{1}{2} \omega_b^2 x^2 .
\end{equation}
The deterministic linearized underdamped dynamics is
\begin{equation}
\begin{gathered}
\dot{x}=v \\
\dot{v}=\omega_b^2 x-\gamma v
\end{gathered}
\end{equation}
Its eigenvalues satisfy
\begin{equation}
\lambda^2+\gamma \lambda-\omega_b^2=0 .
\end{equation}
Thus,
\begin{equation}
\begin{gathered}
\lambda_{+}=\sqrt{\frac{\gamma^2}{4}+\omega_b^2}-\frac{\gamma}{2} \\
\lambda_{-}=-\sqrt{\frac{\gamma^2}{4}+\omega_b^2}-\frac{\gamma}{2}
\end{gathered}
\end{equation}
The unstable reaction coordinate is
\begin{equation}
y=v-\lambda_{-} x .
\end{equation}
In terms of $y$, the local stochastic dynamics near the barrier becomes
\begin{equation}
d y=\lambda_{+} y d t+\sqrt{\frac{2 \gamma}{\beta}} d W_t
\end{equation}
The local committor, i.e. the probability to proceed to the product basin, satisfies
\begin{equation}
\lambda_{+} y q^{\prime}(y)+\frac{\gamma}{\beta} q^{\prime \prime}(y)=0
\end{equation}
with
\begin{equation}
q(-\infty)=0, \quad q(+\infty)=1 .
\end{equation}
The solution is
\begin{equation}
q(y)=\frac{1}{2}\left[1+\operatorname{erf}\left(\sqrt{\frac{\beta \lambda_{+}}{2 \gamma}} y\right)\right] .
\end{equation}
Accordingly, the corrector transition layer has the local form
\begin{equation}
\chi(x, v) \simeq \chi_A+\left(\chi_B-\chi_A\right) q\left(v-\lambda_{-} x\right) .
\end{equation}
This expression shows that the plateau transition is controlled by the phase-space separatrix
\begin{equation}
v-\lambda_{-} x=0,
\end{equation}
not by the naive configurational dividing surface
\begin{equation}
x=0 .
\end{equation}
The corresponding transmission coefficient is
\begin{equation}
\kappa_{\mathrm{GH}}=\frac{\lambda_{+}}{\omega_b},
\end{equation}
which is the Grote-Hynes correction. In the present language, this correction arises because the corrector transition layer is aligned with the unstable phase-space mode rather than with the bare barrier coordinate.

The martingale density makes the same statement in fluctuation form. Since
\begin{equation}
\chi(x, v) \simeq \chi_A+\left(\chi_B-\chi_A\right) q(y), \quad y=v-\lambda_{-} x
\end{equation}
we have
\begin{equation}
\partial_v \chi \simeq\left(\chi_B-\chi_A\right) q^{\prime}(y)
\end{equation}
Using the derivative of the committor,
\begin{equation}
q^{\prime}(y)=\sqrt{\frac{\beta \lambda_{+}}{2 \pi \gamma}} \exp \left[-\frac{\beta \lambda_{+}}{2 \gamma} y^2\right]
\end{equation}
we obtain
\begin{equation}
\partial_v \chi \simeq\left(\chi_B-\chi_A\right) \sqrt{\frac{\beta \lambda_{+}}{2 \pi \gamma}} \exp \left[-\frac{\beta \lambda_{+}}{2 \gamma}\left(v-\lambda_{-} x\right)^2\right] .
\end{equation}
Therefore the underdamped local martingale strength,
\begin{equation}
r_{\mathrm{UD}}(x, v)=\frac{2 \gamma}{\beta}\left[\partial_v \chi(x, v)\right]^2
\end{equation}
is concentrated near
\begin{equation}
v-\lambda_{-} x=0
\end{equation}
This demonstrates that reactive fluctuations are generated not simply at the barrier position, but near the phase-space separatrix selected by the unstable mode.

The underdamped result therefore extends the overdamped picture in two ways. First, the rate remains encoded in a corrector plateau gap. Second, the platform transition and the martingale density become phase-space objects. The zeroth Hermite mode carries the dominant slow-platform response, while the higher velocity modes encode the geometry and fluctuation source of inertial barrier crossing.

{\it Friction dependence and the Kramers–GH–PGH hierarchy.} The plateau-gap construction provides a rate without explicitly invoking a transition-state surface. Nevertheless, in classical double-well dynamics, it must reproduce the known rate hierarchy. We therefore compare
\begin{equation}
k_\chi=\frac{1}{2\left(\chi_R-\chi_L\right)}
\end{equation}
with transition-state theory, the overdamped Kramers asymptote, the Grote-Hynes correction, and the Pollak-Grabert-Hänggi turnover expression.

For the quartic double well,
\begin{equation}
U(x)=\Delta U\left(x^2-1\right)^2,
\end{equation}
the harmonic frequencies at the well bottom and barrier top are
\begin{equation}
\begin{gathered}
\omega_a=\sqrt{U^{\prime \prime}\left(x_a\right)}=\sqrt{8 \Delta U} \\
\omega_b=\sqrt{\left|U^{\prime \prime}\left(x_b\right)\right|}=\sqrt{4 \Delta U}
\end{gathered}
\end{equation}
The transition-state-theory rate is
\begin{equation}
k_{\mathrm{TST}}=\frac{\omega_a}{2 \pi} e^{-\beta \Delta U} .
\end{equation}
This expression counts the equilibrium positive flux through the barrier and assumes that every crossing is reactive. In underdamped dynamics, however, crossing the barrier coordinate does not guarantee escape. The Grote-Hynes correction replaces the bare barrier frequency by the growth rate of the unstable phase-space mode,
\begin{equation}
\lambda_{+}=\sqrt{\frac{\gamma^2}{4}+\omega_b^2}-\frac{\gamma}{2} .
\end{equation}
Thus,
\begin{equation}
k_{\mathrm{GH}}=k_{\mathrm{TST}} \frac{\lambda_{+}}{\omega_b} .
\end{equation}
In the large-friction limit,
\begin{equation}
\lambda_{+} \simeq \frac{\omega_b^2}{\gamma},
\end{equation}
and therefore
\begin{equation}
k_{\mathrm{GH}} \simeq \frac{\omega_a \omega_b}{2 \pi \gamma} e^{-\beta \Delta U},
\end{equation}
which is precisely the overdamped Kramers asymptote.
At low friction, the rate is not controlled only by the barrier-top unstable mode. A trajectory may cross the barrier region many times before losing enough energy to remain in the product basin. The relevant slow process becomes energy diffusion. The PGH expression incorporates this by multiplying the Grote-Hynes rate by a depopulation factor,
\begin{equation}
k_{\mathrm{PGH}}=k_{\mathrm{GH}} \Upsilon(\delta),
\end{equation}
where
\begin{equation}
\delta=\beta \gamma S^{\ddagger}
\end{equation}
is the dimensionless energy loss and $S^{\ddagger}$ is the action evaluated at the barrier energy. The depopulation factor satisfies
\begin{equation}
\Upsilon(\delta) \rightarrow 1, \quad \delta \gg 1,
\end{equation}
and
\begin{equation}
\Upsilon(\delta) \sim \delta, \quad \delta \ll 1 .
\end{equation}
Thus, in the low-friction limit,
\begin{equation}
k_{\mathrm{PGH}} \propto \gamma e^{-\beta \Delta U},
\end{equation}
whereas in the high-friction limit,
\begin{equation}
k_{\mathrm{PGH}} \rightarrow k_{\mathrm{OD}} \propto \gamma^{-1} e^{-\beta \Delta U} .
\end{equation}
This gives the Kramers turnover.

In our formulation, the same hierarchy appears through the corrector platform gap. At large friction, the Hermite-Galerkin solution reduces to the overdamped corrector, and $k_\chi$ approaches the overdamped Kramers rate. At moderate friction, the corrector transition layer is tilted in phase space and aligned with the unstable coordinate $v-\lambda_{-} x$, giving the Grote-Hynes transmission correction. At lower friction, the corrector becomes increasingly influenced by energy-shell dynamics, and the convergence of $k_\chi$ requires higher velocity modes. This reflects the crossover from position-controlled escape to energy-diffusioncontrolled escape. 

The numerical results show that the platform-gap rate follows the classical hierarchy. In the overdamped regime,
\begin{equation}
k_\chi \simeq k_{\mathrm{OD}} .
\end{equation}
In the moderate-friction underdamped regime,
\begin{equation}
k_\chi
\end{equation}
tracks the Grote-Hynes correction. Across the turnover regime, it follows the PGH trend once enough Hermite modes are retained. Therefore, the Kramers, Grote-Hynes, and PGH formulas can be interpreted as different asymptotic manifestations of the same object: the plateau gap of the occupation-time Poisson corrector.

\begin{figure}
    \centering
    \includegraphics[width=1\linewidth]{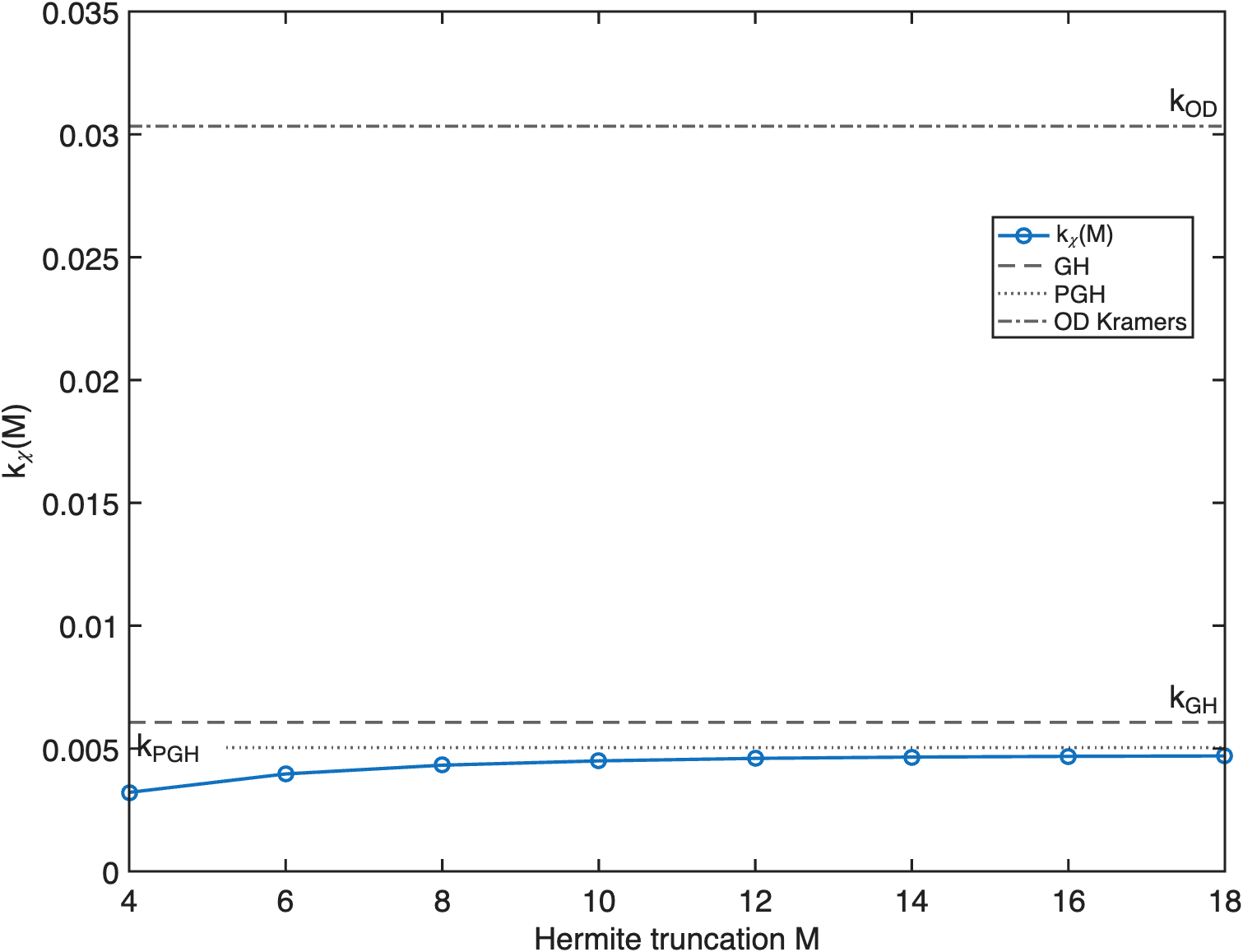}
    \caption{Hermite truncation convergence of the underdamped plateau-gap rate.Rate $k_\chi(M)$ extracted from the corrector plateau gap as a function of the Hermite truncation order $M$. Horizontal lines show the corresponding GH, PGH, and overdamped Kramers reference rates for the same parameters. The plateau-gap rate converges with increasing $M$, demonstrating that the Hermite-Galerkin representation resolves the slow occupation-time corrector. The zeroth mode caries the dominant plateau structure, but convergence of $k_\chi$ requires nonzero velocity modes, which encode inertial recrossing, energy-diffusion corrections, and the martingale channel.}
    \label{fig-fig6}
\end{figure}

This viewpoint also clarifies the role of the Hermite modes. The zeroth coefficient $a_0(x)$ carries the slow occupation response and therefore dominates the plateau gap. However, the nonzero velocity modes renormalize this gap by encoding inertial recrossing and energy-diffusion effects. In the moderate-friction regime, these modes reshape the transition layer into the Grote-Hynes unstable coordinate. In the lowfriction regime, more modes are required because the slow variable is no longer close to position, but approaches the energy
\begin{equation}
E=\frac{1}{2} v^2+U(x) .
\end{equation}
Indeed, applying Itô's formula to the underdamped energy gives

Indeed, applying Itô's formula to the underdamped energy gives
\begin{equation}
d E=-\gamma v^2 d t+\frac{\gamma}{\beta} d t+\sqrt{\frac{2 \gamma}{\beta}} v d W_t
\end{equation}
The last term is an energy martingale. Its quadratic variation,
\begin{equation}
d\left\langle M_E\right\rangle_t=\frac{2 \gamma}{\beta} v^2 d t
\end{equation}
is the stochastic source of energy diffusion. Near the barrier energy, averaging over a nearly Hamiltonian orbit gives an energy loss proportional to
\begin{equation}
\gamma S^{\ddagger} .
\end{equation}
This is the origin of the dimensionless energy-loss parameter
\begin{equation}
\delta=\beta \gamma S^{\ddagger} .
\end{equation}
Thus, in the low-friction limit, the martingale part of the dynamics shifts from a barrier-localized fluctuation channel to an energy-diffusion channel. The PGH depopulation factor can therefore be viewed as the energy-diffusion correction to the corrector platform dynamics.

The important point is not that the Poisson-corrector method replaces the classical formulas in their asymptotic regimes. Rather, it places them in a single rate-extraction framework. The rate is always obtained from the same quantity,
\begin{equation}
\Delta \chi=\chi_B-\chi_A
\end{equation}
while the physical interpretation of this plateau gap changes with friction. In the overdamped regime, it measures the configurational diffusion time across the barrier. In the moderate-friction regime, it incorporates phase-space recrossing. In the low-friction regime, it reflects the slow diffusion of energy required for depopulation of the reactant well.

This unification is useful because it separates two questions that are often mixed in rate theory. The first question is how to extract the scalar rate. The corrector-platform formula answers this through $\Delta \chi$. The second question is what dynamical mechanism generates the reactive fluctuation. The martingale density answers this through
\begin{equation}
r(z)=\nabla \chi(z)^T a(z) \nabla \chi(z) .
\end{equation}
For overdamped dynamics, $r$ identifies the barrier region. For underdamped dynamics, it identifies the phase-space separatrix or, in the low-friction regime, the energy-diffusion channel. The plateau gap gives the rate; the martingale density gives the mechanism.

{\it Nonequilibrium extension and discussion.} The preceding examples use equilibrium double-well dynamics because they provide a direct connection to classical rate theory. However, the corrector-platform construction itself does not rely on equilibrium. It only requires a Markov generator, a stationary measure, and a metastable partition. This is a crucial distinction from Kramers-type theories, which are formulated around a potential barrier, local curvatures, and a thermal equilibrium measure. For a general diffusion process,
\begin{equation}
d Z_t=b\left(Z_t\right) d t+\sigma\left(Z_t\right) d W_t,
\end{equation}
the generator is
\begin{equation}
\mathcal{L}=b(z) \cdot \nabla+\frac{1}{2} a(z): \nabla \nabla, \quad a(z)=\sigma(z) \sigma(z)^T
\end{equation}
If the dynamics admits a stationary distribution $\mu$, one may define the same centered occupation observable,
\begin{equation}
h(z)=\mathbf{1}_B(z)-\mu_B,
\end{equation}
and solve
\begin{equation}
\mathcal{L} \chi=-h, \quad\langle\chi\rangle_\mu=0 .
\end{equation}
Whenever the dynamics has two metastable basins $A$ and $B$, the corrector develops two plateaus. The same plateau-gap formula gives
\begin{equation}
\begin{aligned}
k_{A \rightarrow B} & =\frac{\mu_B}{\chi_B-\chi_A} \\
k_{B \rightarrow A} & =\frac{\mu_A}{\chi_B-\chi_A}
\end{aligned}
\end{equation}
No detailed-balance condition is used in this argument. The rates follow from the coarse-grained twostate generator and the Poisson equation for the occupation-time observable.

This observation is important for nonequilibrium metastability. In an irreversible steady state, the drift field may contain rotational or nonconservative components, and the stationary probability current may be nonzero. Then there may be no scalar potential $U$, no Boltzmann weight $e^{-\beta U}$, and no obvious barrier top. Transition-state constructions become ambiguous because the location of a dividing surface may depend on the chosen coordinates or on the dominant reactive current. In contrast, the corrector-platform method asks a different question: starting from a point $z$, what is the integrated future bias of the product-basin occupation? This response is precisely the Poisson corrector.

The plateau gap therefore plays the role of a dynamical response time rather than a geometric barrier height. If $z$ lies deep in basin $A$, the future occupation of $B$ is initially suppressed, and $\chi(z)$ approaches the $A$-plateau. If $z$ lies deep in $B$, the future occupation of $B$ is enhanced, and $\chi(z)$ approaches the $B$ plateau. The time required for the process to lose memory of these two initial conditions is encoded in the plateau separation. This interpretation remains valid whether the underlying dynamics is reversible or irreversible.

The martingale density provides the complementary nonequilibrium diagnostic. For the general diffusion above, the martingale part of the occupation-time decomposition is
\begin{equation}
M_t=\int_0^t \nabla \chi\left(Z_s\right)^T \sigma\left(Z_s\right) d W_s
\end{equation}
with quadratic variation
\begin{equation}
\langle M\rangle_t=\int_0^t \nabla \chi\left(Z_s\right)^T a\left(Z_s\right) \nabla \chi\left(Z_s\right) d s
\end{equation}
Thus,
\begin{equation}
r(z)=\nabla \chi(z)^T a(z) \nabla \chi(z)
\end{equation}
is the local reactive-noise strength. The stationary weighted field,
\begin{equation}
R(z)=r(z) \mu(z),
\end{equation}
identifies where stochastic forcing contributes most strongly to long-time occupation fluctuations. In equilibrium overdamped dynamics, this field localizes near the barrier. In underdamped dynamics, it localizes near a phase-space separatrix or an energy-diffusion channel. In nonequilibrium systems, it can reveal reactive-noise hotspots that are not associated with any potential barrier.

This is the central advantage of the formulation. A scalar rate answers how fast a reaction occurs; the martingale density answers where the stochastic part of that reaction is generated. These two pieces of information come from the same Poisson problem. The corrector $\chi$ determines the slow response and the platform gap; its gradient, weighted by the diffusion matrix, determines the fluctuation source.

The method also clarifies the relation between rate extraction and committor-based descriptions. The committor solves a homogeneous boundary-value problem and gives the probability of reaching $B$ before $A$. It identifies reactive geometry. The occupation-time corrector solves an inhomogeneous Poisson equation and gives the integrated future occupation bias. It identifies the time scale of the slow mode. In the high-barrier limit, the corrector transition layer is often controlled by the committor, but the two objects are not identical. The committor answers where a trajectory is likely to go; the corrector answers how long the occupation imbalance persists. This is why the plateau gap carries rate information.

The approach also differs from direct mean-first-passage-time calculations. A mean first-passage time depends on a chosen initial distribution and target boundary. The corrector-platform formula instead extracts both forward and backward rates from a stationary occupation observable. For two-state metastability, the stationary weights $\mu_A, \mu_B$ and the single plateau gap $\Delta \chi$ determine the two directional rates. This makes the method naturally compatible with coarse-graining and with data-driven approximations of $\mathcal{L}$ or $\chi$.

Several numerical routes are possible. In low-dimensional systems, $\chi$ can be obtained by finite differences, finite elements, or spectral expansions. For underdamped dynamics, Hermite-Galerkin expansion in the velocity coordinate provides a natural representation. For high-dimensional systems, one may approximate the corrector from trajectories using Galerkin regression, neural trial functions, or operator-learning methods. Once $\chi$ is learned, both the rate and the martingale density follow directly.

The present work focuses on double-well dynamics because it permits controlled comparison with Kramers, Grote-Hynes, and PGH theories. In this setting, the corrector-platform method does not replace the classical asymptotic formulas; rather, it reorganizes them. The Kramers rate appears as an overdamped plateau gap. The Grote-Hynes correction appears as the phase-space tilt of the corrector transition layer. The PGH depopulation correction appears as the low-friction energy-diffusion modification of the platform dynamics. The same Poisson problem therefore contains the rate, the recrossing geometry, and the fluctuation source.

This perspective suggests a practical criterion for metastability. If $\chi$ does not form well-defined plateaus, the proposed two-state coarse-graining is not dynamically justified. Conversely, clear plateaus indicate that a two-state rate description is meaningful, and their separation gives the corresponding rates. Thus, the corrector is not only a rate estimator but also a diagnostic of whether a metastable partition is appropriate.

{\it Conclusion.} We have shown that metastable reaction rates can be extracted from the plateau gap of an occupationtime Poisson corrector. For two metastable basins $A$ and $B$, the corrector solving

\begin{equation}
\mathcal{L} \chi=-\left(\mathbf{1}_B-\mu_B\right)
\end{equation}

forms two basin plateaus, and their separation gives

\begin{equation}
k_{A \rightarrow B}=\frac{\mu_B}{\chi_B-\chi_A}, \quad k_{B \rightarrow A}=\frac{\mu_A}{\chi_B-\chi_A} .
\end{equation}

This relation turns reaction-rate extraction into a plateau-gap problem.
The associated martingale term provides a second field-level quantity. Its quadratic variation defines

\begin{equation}
r(z)=\nabla \chi(z)^T a(z) \nabla \chi(z),
\end{equation}

which identifies where stochastic forcing generates reactive fluctuations. In overdamped dynamics, this reactive-noise density localizes near the barrier. In underdamped dynamics, it localizes in phase space near the unstable separatrix or, in the low-friction regime, along the energy-diffusion channel.

For double-well dynamics, the corrector-platform construction recovers the Kramers, Grote-Hynes, and PGH hierarchy. The scalar rate is encoded in the corrector plateau gap, while the physical mechanism of the transition is encoded in the martingale density. Because the method requires only a generator, a stationary measure, and a metastable partition, it provides a route to reaction-rate extraction beyond equilibrium barrier-crossing theory. This framework reframes metastable reactions as a problem of slow occupation response: the rate is read from the corrector, and the reactive noise is read from its martingale.

\bibliographystyle{IEEEtran}
\bibliography{ref.bib}
\end{document}